\documentstyle[12pt,a4]{article}
\newcommand{\beq}{\begin{equation}}
\newcommand{\eeq}{\end{equation}}
\newcommand{\bea}{\begin{eqnarray}}
\newcommand{\eea}{\end{eqnarray}}
\newlength{\bredde}
\def\slash#1{\settowidth{\bredde}{$#1$}\ifmmode\,\raisebox{.15ex}{/}
\hspace*{-\bredde} #1\else$\,\raisebox{.15ex}{/}\hspace*{-\bredde} #1$\fi}
\begin{document}
\title{Effective Field Theories from QCD\thanks{Based on work done in
collaboration with P.H.~Damgaard and H.B.~Nielsen.}}
\author{R. Sollacher\\ GSI, P.O.Box 110552, D-64220 Darmstadt, Germany}
\maketitle
\begin{abstract}
We present a method for extracting effective Lagrangians from QCD.
The resulting effective Lagrangians are based on exact rewrites of
cut-off QCD in terms of these new collective field degrees of freedom.
These cut-off Lagrangians are thus ``effective'' in the sense that
they explicitly contain some of the physical long-distance degrees of
freedom from the outset. As an example we discuss the introduction of
a new collective field carrying the quantum numbers of the
$\eta'$-meson.
\end{abstract}
\vskip1cm
\noindent{\em Contribution presented by R. Sollacher at the workshop
``QCD'94'', Montpellier, France, July 7-13, 1994. To appear in those
proceedings.}

\section{Introduction}

One possible test for low-energy QCD would be
a direct calculation of the parameters of chiral perturbation theory.
Of course, we can not provide the solution of such an ambitious
program. However, we will demonstrate a method for doing the first
step in this direction. We show how to introduce the appropriate
effective degrees of freedom and how to extract a recipe for
determining the parameters of a suitable effective Lagrangian.
For technical reasons we concentrate on the simpler Abelian case of
an effective field carrying the quantum numbers of a flavour singlet
pseudoscalar meson. Such an effective field should describe the
dynamics of the $\eta'$-meson. Finally, we outline the extension of
this method to the more interesting non-Abelian case resulting in an
effective Lagrangian analogous to the one of chiral perturbation
theory.

Our main tool is the {\em gauge-symmetric collective field technique}
\cite{HoKi,AlfDam:90}. It was used to show that
(1+1)-dimen\-sio\-nal {\em
bosonization} (and fermionization) \cite{boson} are only two extremes
of a continuum of equivalent field theory descriptions
\cite{AlfDam:90,us}. In fact, both
can be viewed as particular {\em gauge fixings} of a ``higher''
gauge-symmetric theory that contains both bosons and fermions \cite{us}.
The extent to which this method can be applied to QCD has recently
been discussed in ref. \cite{us1}, and the purpose of this talk is to
provide a short review of that work.

\section{The $\eta'$-meson from QCD}

\subsection{From QCD to ``super''-QCD}

Our starting point is a generating functional for QCD with $N_f$
flavours in the chiral limit of vanishing quark masses. In Euclidean
space-time it is of the form
\begin{eqnarray}
{\cal Z}_{QCD} [A] &=& \int\! {\cal D} [\bar{\psi},\psi]{\cal
D}\mu[G]\; e^{-\int\! d^4x\; {\cal L}_{QCD}} \cr
{\cal L}_{QCD} &=& \bar{\psi}(x) (\slash{D}
-i \slash{A} (x)\gamma_5)\psi(x) \cr
&& +\frac{1}{4g^2} tr G_{\mu\nu} (x)G_{\mu\nu} (x)~.
\label{eq:ZQCD}
\end{eqnarray}
The external Abelian axial vector source $A_\mu(x)$ serves to define
appropriate Green functions through functional differentiation. The
$SU(N_c)$-valued covariant derivative is denoted by $D_\mu$ and
$G_{\mu\nu}(x)$ is the corresponding field strength tensor.
The gauge-fixing terms of the usual $SU(N_c)$ colour gauge symmetry
are implicitly part of the gluon measure ${\cal{D}}\mu[G]$.
For the following we only have to specify explicitly a
regularization for the fermionic sector. A convenient
consistent scheme in this sector is provided by a set of Pauli-Villars
regulator fields \cite{Ball}.

As the field transformation introducing a pseudoscalar field $\theta
(x)$ we choose a chiral rotation of the quark fields:
\beq
\psi (x) = e^{i\theta (x) \gamma_5} \chi (x)~~,~~\bar{\psi} (x) =
\bar{\chi} (x) e^{i\theta (x) \gamma_5}
\label{eq:trans}
\eeq
In order to promote $\theta (x)$ into a
dynamical field we integrate over all possible configurations.
The result is the generating functional of what one may call
``super''-QCD:
\begin{eqnarray}
{\cal Z}_{SQCD} [A] &=& \int\! {\cal D}[\theta ]{\cal D}
[\bar{\chi},\chi ]{\cal D}\mu[G]\; e^{-\int\! d^4x\; {\cal L}_{SQCD}}
\cr
{\cal L}_{SQCD} &=& {\cal L}_{QCD} + i\bar{\chi}
\slash{\partial}\theta \gamma_5 \chi \cr
&& +\frac{N_f}{2} \partial_\mu \theta f^2 \partial_\mu \theta - N_f A_\mu
f^2 \partial_\mu \theta \cr
&& -2iN_f \theta Q + \ldots~.
\label{eq:Znew}
\end{eqnarray}
The terms in the last two lines arise from the Jacobian of the
transformation \cite{Ball}. Only the leading terms of an expansion in
powers of $\partial_\mu\theta$ and $A_\mu$ are shown. The term in the
last line is due to the $U(1)$-anomaly, with the instanton density
\beq
Q = \frac{1}{32 \pi^2}\epsilon_{\mu\nu\rho\sigma} tr G_{\mu\nu}
G_{\rho\sigma}  ~~.
\eeq
Assuming that the gluonic measure implies a
summation over all integer instanton numbers $N_{inst}$ one immediately
realizes that $\theta (x)$ is globally periodic, $i.e.$,
\beq
\theta (x) \equiv \theta (x) + \frac{n \pi}{N_f}~~.
\eeq
Thus, one can restrict the $\theta$-integration globally to the
interval $[0,\pi/N_f ]$.

The object $f^2$ in eq. (\ref{eq:Znew}) is a nonlocal operator:
\bea
f^2 &=& -\frac{N_c \kappa_2 \Lambda^2}{2\pi^2} + \frac{N_c}{12\pi^2}
\partial^2 - \frac{N_c \kappa_{-2}}{24\pi^2 \Lambda^2} \partial^2
\partial^2\cr
&& - \frac{\kappa_{-2}}{24 \pi^2
\Lambda^2} \; tr_c  G_{\nu\rho} G_{\nu\rho} + \ldots
\label{eq:f2}
\eea
For simplicity we have displayed the leading terms of a gradient
expansion. The coefficients $\kappa_2, \kappa_{-2}$ in eq.(\ref{eq:f2})
are regularization-scheme dependent constants \cite{us1}.
There are two important features of $f^2$:
\begin{itemize}
\item It induces a higher-derivative (or
essentially Pauli-Villars) {\em regularized} bosonic propagator with a
regulator mass proportional to $\Lambda^2$, at least in a perturbative
sense.
\item The possible occurrence of gluonic condensates in (\ref{eq:f2})
signals spontaneous chiral symmetry breaking.
However, in the presence of gluonic condensates the implied
expansion in inverse powers of the cutoff may fail to converge and one
has to live with the full nonlocal structure of $f^2$.
\end{itemize}

Finally, as a consequence of the integration over the collective
field $\theta (x)$ a chiral gauge symmetry appears \cite{AlfDam:90}:
\bea
\chi (x) &\to& e^{i\alpha (x) \gamma_5} \chi (x) \cr
\bar{\chi} (x) &\to& \bar{\chi} (x) e^{i\alpha (x) \gamma_5} \cr
\theta (x) &\to& \theta (x) - \alpha (x).
\label{eq:sym5}
\eea
It is this gauge symmetry which has to be removed by suitable gauge
fixing terms in order to leave the generating functional $Z_{QCD} [A]$
unaltered.

\subsection{A partial bosonization}

For that purpose we consider the change in the divergence of the axial
singlet current:
\beq
i\partial_\mu \langle \bar{\psi} \gamma_\mu \gamma_5 \psi \rangle =
i\partial_\mu \langle \bar{\chi} \gamma_\mu \gamma_5 \chi \rangle +
N_f \partial_\mu \langle f^2 \partial_\mu \theta \rangle + \ldots
\label{eq:trAxCur}
\eeq
The additional terms represented by dots are at least of third order
in $\theta(x)$. In order to saturate the anomalous chiral Ward identities by
the field $\theta (x)$ we have to eliminate the $\chi$-dependent part
of eq. (\ref{eq:trAxCur}). The resulting gauge fixing function $\Phi
(x)$ reads:
\beq
\Phi =  i\frac{\partial_\mu}{N_f f_0^2\partial^2}
\bar{\chi} \gamma_\mu \gamma_5\chi ~.
\label{eq:Phi'}
\eeq
For simplicity we have replaced $f^2$ by a constant $f_0^2$ to be
explained later on. The function $\Phi$ is defined in a very formal
manner on account of the inverse Laplacian. Its presence ensures the
proper gauge fixing of global chiral gauge transformations.

We now follow the standard Faddeev-Popov
procedure introducing a field $b$ ensuring the gauge constraint
$\Phi (x) =0$ as well as Grassmannian ghosts $\bar{c}$ and $c$. The
original generating functional $Z_{QCD}[A]$ now appears
as
\begin{eqnarray}
{\cal Z}_{QCD} [A] &=& \int\! {\cal D}[\theta ]{\cal
D}[\bar{\chi},\chi ]{\cal D}\mu[G] {\cal D} [b,c,\bar{c}] \cr
&&\qquad e^{-\int\! d^4x\; {\cal L}'}
\cr
{\cal L}' &=& {\cal L}_{SQCD} + \frac{i}{N_f f_0^2} \bar{\chi}
\slash{B} \gamma_5 \chi + \bar{c} c + \ldots
\label{eq:Zgf}
\end{eqnarray}
The field $b$ is related to the purely longitudinal axial vector field
$B_\mu$ by the relation
\beq
b(x) = \partial_\mu B_\mu (x)~~.
\eeq
The global periodicity of the chiral gauge transformation implies that
the measure for $b$ extends over an infinite number of topological
sectors:
\beq
\int\! {\cal D}[b] = \sum_{k=-\infty}^{+\infty} \int\! {\cal D} [b]_k
\eeq
For each integer $k$ the field $b$ obeys the topological constraint
\beq
\int\! d^4x\; b(x) = \int\! d^4x\; \partial_\mu B_\mu (x) = ikN_f~~.
\label{eq:top}
\eeq
The dots in ${\cal L}'$ in (\ref{eq:Zgf}) indicate higher order terms
in $b$ which are necessary due to the fact that the gauge fixing term
itself modifies the regularized fermionic measure.

\subsection{The physical content of $\theta (x)$}

The field $\theta (x)$ saturates the chiral Ward identities for
the divergence of the axial singlet current by construction. The
spectrum of this operator, which includes the $\eta'$, is now
described by $\theta$.
Just in order to illustrate how non-trivial results can be extracted from
the effective Lagrangian ${\cal L}'$ in  (\ref{eq:Zgf}), let us
integrate out all fields in (\ref{eq:Zgf}) except
$\theta$ to arrive at an effective Lagrangian
\beq
{\cal L}_{eff} = \frac{F^2_0}{2} \partial_\mu \theta
\partial_\mu \theta +
\frac{F_0^2 M_0^2}{2} \theta^2 + \ldots~~.
\label{eq:Leff}
\eeq
The dots denote higher derivative terms and self-interactions of order
$\theta^3$. The parameters $F_0$ and $M_0$ are defined through
\beq
F_0^2 M_0^2 = 4 N_f^2 \int\! d^4x\; \left\langle Q(x)Q(0)
\right\rangle_{trunc}
\label{eq:M0}
\eeq
and
\beq
F_0^2 = N_f f_0^2 - \frac{N_f^2}{2} \int\! d^4x\; x^2 \left\langle
Q(x)Q(0) \right\rangle_{trunc}~~,
\label{eq:F0}
\eeq
where $f_0^2 $ is just the zero-momentum limit of $\langle f^2
\rangle_{trunc}$.

The identification of $F_0$ and $M_0$ with the decay constant and mass
of the $\eta'$-meson can be made only in the case where this meson is
substantially lighter than other excitations with the same quantum
numbers. This is the case in the limit $N_c \to \infty$
where ({\ref{eq:M0}) reduces to the relation derived by Witten
\cite{Witten} and Veneziano \cite{Veneziano}. However, the expectation
values $\langle \ldots \rangle_{trunc}$ have to be taken with respect
to a ``truncated'' version of QCD:
\beq
{\cal L}_{trunc} = {\cal L}_{SQCD} + \frac{i}{N_f f_0^2} \bar{\chi}
\slash{B} \gamma_5 \chi + \bar{c} c + \ldots
\label{eq:Ltrunc}
\eeq
Witten \cite{Witten} argued that in the large-$N_c$ limit the mass of
the $\eta'$ can be derived from the topological susceptibility of pure
Yang-Mills theory, $i.e.$ QCD without quarks. Here, we are not
removing the quarks completely. It is the topological property of the
Lagrange-multiplier field $b$ which provides a nonvanishing
topological susceptibility and thus a mass for the $\eta'$. The reason
is that fermionic zero modes in the presence of topological nontrivial
fields imply a constraint
\beq
N_{inst} + k = 0
\eeq
As we sum over $k$ we also cover all instanton sectors.

We could go beyond the large-$N_c$ limit if we skip the gradient
expansion leading to the effective Lagrangian (\ref{eq:Leff}). Then we
can still derive the propagator for $\theta$ containing contributions
from all possible excitations with the quantum numbers of the $\eta'$.

\section{The non-Abelian case: An outline}

Effective fields for the $SU(N_f)$ pseudoscalar multiplet can be
introduced along similar lines. After a suitable chiral transformation
one has to fix a gauge. One can again use the saturation of chiral Ward
identities by the new fields as the guiding demand. Integrating out
all fields except the pseudoscalars followed by a gradient expansion
should result in an effective Lagrangian analogous to the one of
chiral perturbation theory. Within such an approach one could
determine relations between the parameters of the resulting effective
Lagrangian. An interesting alternative would be the application of
this approach to lattice-QCD. The calculation of the parameters of
chiral perturbation theory could then be done numerically. This would
require relatively small lattices due to the absence of
Goldstone-bosons in the truncated theory.

\bibliographystyle{unsrt}

\end{document}